\documentclass[twoside]{article}
\usepackage{graphicx}
\usepackage{amsmath}

\catcode`\@=11
\long\def\@makefntext#1{
\protect\noindent \hbox to 3.2pt {\hskip-.9pt
$^{{\eightrm\@thefnmark}}$\hfil}#1\hfill}		

\def\@makefnmark{\hbox to 0pt{$^{\@thefnmark}$\hss}}	

\def\ps@myheadings{\let\@mkboth\@gobbletwo
\def\@oddhead{\hbox{}
\rightmark\hfil\eightrm\thepage}
\def\@oddfoot{}\def\@evenhead{\eightrm\thepage\hfil
\leftmark\hbox{}}\def\@evenfoot{}
\def\sectionmark##1{}\def\subsectionmark##1{}}

\oddsidemargin=\evensidemargin
\newcounter{sectionc}\newcounter{subsectionc}\newcounter{subsubsectionc}
\renewcommand{\section}[1] {\vspace{12pt}\addtocounter{sectionc}{1}
\setcounter{subsectionc}{0}\setcounter{subsubsectionc}{0}\noindent
	{\tenbf\thesectionc. #1}\par\vspace{5pt}}
\renewcommand{\subsection}[1] {\vspace{12pt}\addtocounter{subsectionc}{1}
	\setcounter{subsubsectionc}{0}\noindent
	{\bf\thesectionc.\thesubsectionc. {\kern1pt \bfit #1}}\par\vspace{5pt}}
\renewcommand{\subsubsection}[1] {\vspace{12pt}\addtocounter{subsubsectionc}{1}
	\noindent{\tenrm\thesectionc.\thesubsectionc.\thesubsubsectionc.
	{\kern1pt \tenit #1}}\par\vspace{5pt}}
\newcommand{\nonumsection}[1] {\vspace{12pt}\noindent{\tenbf #1}
	\par\vspace{5pt}}

\newcounter{appendixc}
\newcounter{subappendixc}[appendixc]
\newcounter{subsubappendixc}[subappendixc]
\renewcommand{\thesubappendixc}{\Alph{appendixc}.\arabic{subappendixc}}
\renewcommand{\thesubsubappendixc}
	{\Alph{appendixc}.\arabic{subappendixc}.\arabic{subsubappendixc}}

\renewcommand{\appendix}[1] {\vspace{12pt}
        \refstepcounter{appendixc}
        \setcounter{figure}{0}
        \setcounter{table}{0}
        \setcounter{lemma}{0}
        \setcounter{theorem}{0}
        \setcounter{corollary}{0}
        \setcounter{definition}{0}
        \setcounter{equation}{0}
        \renewcommand{\thefigure}{\Alph{appendixc}.\arabic{figure}}
        \renewcommand{\thetable}{\Alph{appendixc}.\arabic{table}}
        \renewcommand{\theappendixc}{\Alph{appendixc}}
        \renewcommand{\thelemma}{\Alph{appendixc}.\arabic{lemma}}
        \renewcommand{\thetheorem}{\Alph{appendixc}.\arabic{theorem}}
        \renewcommand{\thedefinition}{\Alph{appendixc}.\arabic{definition}}
        \renewcommand{\thecorollary}{\Alph{appendixc}.\arabic{corollary}}
        \renewcommand{\theequation}{\Alph{appendixc}.\arabic{equation}}
        \noindent{\tenbf Appendix \theappendixc #1}\par\vspace{5pt}}
\newcommand{\subappendix}[1] {\vspace{12pt}
        \refstepcounter{subappendixc}
        \noindent{\bf Appendix \thesubappendixc. {\kern1pt \bfit #1}}
	\par\vspace{5pt}}
\newcommand{\subsubappendix}[1] {\vspace{12pt}
        \refstepcounter{subsubappendixc}
        \noindent{\rm Appendix \thesubsubappendixc. {\kern1pt \tenit #1}}
	\par\vspace{5pt}}

\newcommand{\textlineskip}{\baselineskip=13pt}
\newcommand{\smalllineskip}{\baselineskip=10pt}

\def\eightcirc{
\begin{picture}(0,0)
\put(4.4,1.8){\circle{6.5}}
\end{picture}}
\def\eightcopyright{\eightcirc\kern2.7pt\hbox{\eightrm c}}

\newcommand{\copyrightheading}[1]
	{\vspace*{-2.5cm}\smalllineskip{\flushleft
	{\footnotesize International Journal of Modern Physics C, #1}\\
	{\footnotesize $\eightcopyright$\, World Scientific Publishing
	 Company}\\
	 }}

\newcommand{\publisher}[2]{{\begin{center}\footnotesize\smalllineskip
	Received #1\\
	Revised #2
	\end{center}
	}}

\def\abstracts#1#2#3{{
	\centering{\begin{minipage}{4.5in}\baselineskip=10pt\footnotesize
	\parindent=0pt #1\par
	\parindent=15pt #2\par
	\parindent=15pt #3
	\end{minipage}}\par}}

\def\keywords#1{{
	\centering{\begin{minipage}{4.5in}\baselineskip=10pt\footnotesize
	{\footnotesize\it Keywords}\/: #1
	\end{minipage}}\par}}

\renewenvironment{thebibliography}[1]
        {\frenchspacing
	 \ninerm\baselineskip=11pt
         \begin{list}{\arabic{enumi}.}
        {\usecounter{enumi}\setlength{\parsep}{0pt}
	 \setlength{\leftmargin 12.7pt}{\rightmargin 0pt} 
         \setlength{\itemsep}{0pt} \settowidth
	{\labelwidth}{#1.}\sloppy}}{\end{list}}

\newcounter{itemlistc}
\newcounter{romanlistc}
\newcounter{alphlistc}
\newcounter{arabiclistc}

\newcommand{\fcaption}[1]{
        \refstepcounter{figure}
        \setbox\@tempboxa = \hbox{\footnotesize Figure~\thefigure. #1}
        \ifdim \wd\@tempboxa > 5in
           {\begin{center}
        \parbox{5in}{\footnotesize\smalllineskip Figure~\thefigure. #1}
            \end{center}}
        \else
             {\begin{center}
             {\footnotesize Figure~\thefigure. #1}
              \end{center}}
        \fi}

\newcommand{\tcaption}[1]{
        \refstepcounter{table}
        \setbox\@tempboxa = \hbox{\footnotesize Table~\thetable. #1}
        \ifdim \wd\@tempboxa > 5in
           {\begin{center}
        \parbox{5in}{\footnotesize\smalllineskip Table~\thetable. #1}
            \end{center}}
        \else
             {\begin{center}
             {\footnotesize Table~\thetable. #1}
              \end{center}}
        \fi}

\def\@citex[#1]#2{\if@filesw\immediate\write\@auxout
	{\string\citation{#2}}\fi
\def\@citea{}\@cite{\@for\@citeb:=#2\do
	{\@citea\def\@citea{,}\@ifundefined
	{b@\@citeb}{{\bf ?}\@warning
	{Citation `\@citeb' on page \thepage \space undefined}}
	{\csname b@\@citeb\endcsname}}}{#1}}

\newif\if@cghi
\def\cite{\@cghitrue\@ifnextchar [{\@tempswatrue
	\@citex}{\@tempswafalse\@citex[]}}
\def\citelow{\@cghifalse\@ifnextchar [{\@tempswatrue
	\@citex}{\@tempswafalse\@citex[]}}
\def\@cite#1#2{{$\null^{#1}$\if@tempswa\typeout
	{IJCGA warning: optional citation argument
	ignored: `#2'} \fi}}
\newcommand{\citeup}{\cite}

\def\pmb#1{\setbox0=\hbox{#1}
	\kern-.025em\copy0\kern-\wd0
	\kern.05em\copy0\kern-\wd0
	\kern-.025em\raise.0433em\box0}

\def\fnt#1#2{\footnotetext{\kern-.3em
	{$^{\mbox{\scriptsize #1}}$}{#2}}}

\def\fpage#1{\begingroup
\voffset=.3in
\thispagestyle{empty}\begin{table}[b]\centerline{\footnotesize #1}
	\end{table}\endgroup}

\def\runninghead#1#2{\pagestyle{myheadings}
\markboth{{\protect\footnotesize\it{\quad #1}}\hfill}
{\hfill{\protect\footnotesize\it{#2\quad}}}}
\font\tenrm=cmr10
\font\tenit=cmti10
\font\tenbf=cmbx10
\font\bfit=cmbxti10 at 10pt
\font\ninerm=cmr9

\font\eightrm=cmr8

\def\qed{\hbox{${\vcenter{\vbox{			
   \hrule height 0.4pt\hbox{\vrule width 0.4pt height 6pt
   \kern5pt\vrule width 0.4pt}\hrule height 0.4pt}}}$}}

\def\bsc{{\sc a\kern-6.4pt\sc a\kern-6.4pt\sc a}}  	
\def\bflatex{\bf L\kern-.30em\raise.3ex\hbox{\bsc}\kern-.14em 
T\kern-.1667em\lower.7ex\hbox{E}\kern-.125em X}

\begin{document}
\runninghead{K.~Malarz \& A.~Z.~Maksymowicz}{A simple solid-on-solid model of epitaxial films growth: surface roughness and dynamics}
\normalsize\textlineskip
\thispagestyle{empty}
\setcounter{page}{1}
\copyrightheading {Vol. 10, No. 0 (1999) 000--000}
\vspace*{0.88truein}
\fpage{1}

\centerline{\bf A SIMPLE SOLID-ON-SOLID MODEL}
\centerline{\bf OF EPITAXIAL FILMS GROWTH:}
\centerline{\bf SURFACE ROUGHNESS AND DYNAMICS}
\vspace*{0.37truein}
\centerline{\footnotesize K.~MALARZ$^*$ and A.~Z.~MAKSYMOWICZ$^\dag$}
\vspace*{0.015truein}
\centerline{\footnotesize\it Department of Theoretical and Computational Physics,}
\centerline{\footnotesize\it Faculty of Physics and Nuclear Techniques, University of Mining and Metallurgy (AGH)}
\centerline{\footnotesize\it al. Mickiewicza 30, PL-30059 Krak\'ow, Poland}
\centerline{\footnotesize\it E-mail: $^*${\tt malarz@agh.edu.pl}, $^\dag${\tt amax@agh.edu.pl}}
\vspace*{0.225truein}
\publisher{(received date)}{(revised date)}
\vspace*{0.21truein}

\abstracts{The random deposition model must be enriched to reflect the variety of surface roughness due to some material characteristics of the film growing by vacuum deposition or sputtering.
The essence of the computer simulation in this case is to account for possible surface migration of atoms just after the deposition, in connection with binding energy between atoms (as the mechanism provoking the diffusion) and/or diffusion energy barrier.
The interplay of these two factors leads to different morphologies of the growing surfaces from flat and smooth ones, to rough and spiky ones.
In this paper we extended our earlier calculation by applying some extra
diffusion barrier at the edges of terrace-like structures, known as
Ehrlich-Schwoebel barrier. It is experimentally observed that atoms avoid
descending when the terrace edge is approach and these barriers mimic
this tendency. Results of our Monte Carlo computer simulations are discussed
in terms of surface roughness, and compared with other model calculations and some experiments from literature.
The power law of the surface roughness $\sigma$ against film thickness $t$ was confirmed.
The nonzero minimum value of the growth exponent $\beta$ near $0.2$ was obtained which is due to the limited range of the surface diffusion and the Ehrlich-Schwoebel barrier.
Observations for different diffusion range are also discussed.
The results are also confronted with some deterministic growth models.}{}{}
\vspace*{10pt}
\vspace*{1pt}\textlineskip

\vspace*{10pt}

\section{Introduction}
\vspace*{-0.5pt}
\noindent
The surface growth phenomena are of interest for scientists because of thin films numerous industrial aspects as well as examples of that phenomena in crystal growth, biology, {\it etc}.\citeup{herrmann86,gouyet91}
A great variety of models, approaches and approximations are devoted to this problem, among them many theories of solid film surfaces growth based on continuous as well as discrete models.\citeup{levi97,kotrla92a}
The so called solid-on-solid (SOS) model is usually assumed.
In this model no pores or voids or even overhangs are considered and each particle sits on the top of other one.
In SOS model the film surface morphology is described by single-valued function $h(\vec{r},t)$ of surface height in planar coordinate $\vec{r}$ at time $t$.
However, such information is usually too rich and not necessary.
Instead, we characterize the surface in terms of some statistical parameters describing surface roughness, spatial correlations, anisotropy or growth dynamics.
In Ref.~\cite{maksymowicz96} a model for (1+1)-dimensional case is presented.
We use this model for the (2+1)-dimensional case above the percolation limit when the substrate is covered by at least one monolayer (ML) on average.
We study surface morphology and time evolution of the surface width (spatial height-height auto-correlations).
The results of simulations are compared with theoretical predictions as well as accessible experimental data.

\section{Model}
\vspace*{-0.5pt}
\noindent
In the presented model particles simply fall down until they reach the top of a column at randomly chosen position $\vec{r}=(x,y)$ on a two-dimensional lattice.
Each particle is represented by a unit-volume cube which can occupy only discrete position in the lattice.
The simple cubic symmetry is assumed for simplicity of computations.

	To mimic molecular beam epitaxy (MBE) growth, one assumes that the dominant physical mechanism of the surface smoothing (or roughening) is surface diffusion.
We assume that desorption from the surface can be neglected.
We also simplify the situation to the SOS model where no overhangs or bulk vacancies are created.
All these assumption are reasonably well satisfied in typical experimental situations.

	After initial contact to the surface, the particle is allowed to make several one step jumps to the nearest neighbors.
We allow for $L_{\text{dif}}$ jumps which defines the diffusion range.
	The probability of a particle to settle at a given site is given by the Boltzmann factor: $p\propto\exp(-E/kT)$, where $E$ is particle energy at the site, $k$ is the Boltzmann constant and $T$ denotes the temperature.
There are three contributions to the energy $E$.
The first one is the particle-particle binding energy $J$ between the nearest neighbors.
Then the diffusion barrier $V$ enters if we claim that the particle jumps to a neighboring site.
The third contribution $S$ is an interaction energy between the top particle and underlying atom.
The $S$-term is an invention used in simulations to account for the edge effect.
It was shown\citeup{ehrlich66,schwoebel66,schwoebel69} that atoms at the edge of a terrace are reluctant to step down.
This situation was simulated in theoretical study by an additional energy barrier for the edge atoms.\citeup{markov96,markov97}
This barrier is known in the literature as the Ehrlich-Schwoebel (ES) barrier, although its origin is still not quite comprehended in terms of a physical mechanism directly responsible for such a barrier.
Note, that in our model the ES-like barrier does not make it difficult to hop down the step edge, but only to approach the step-edge site (similarly like in Ref.~\cite{smilauer93b} for instance).

\begin{figure}
\centering
\includegraphics[width=95mm]{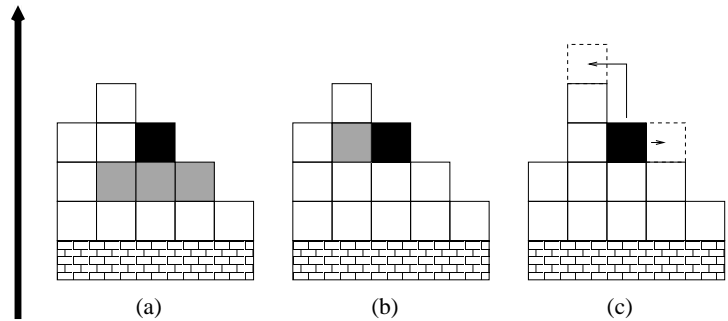}
\caption{A cross-section through the film growing on the substrate.
Squares denote actually occupied sites in the lattice.
The black squares are new arriving particles.
(a) The gray squares are atoms which contribute to the $S$-term.
(b) The gray square is atom which contributes to $J$.
(c) If the new (black) atom jumps to the nearest site, it must overcome the diffusion barrier $V$.
The line of bricks indicates substrate and fat arrow shows direction of growth.}
\label{fig_model_1}
\end{figure}

Finally we get
\begin{equation}
p(\text{stay}) \propto \exp\frac{-(\sum J+\sum S)}{kT},
\label{probstay_eq}
\end{equation}
and
\begin{equation}
p(\text{move left, right, forward, backward}) \propto \exp\frac{-(V+\sum J+\sum S)}{kT},
\label{probmove_eq}
\end{equation}
where the summation in $\sum J$ is over nearest neighbors, and the sum $\sum S$ runs over nearest and next-nearest neighbors of the underlying layer, as shown in Fig.~\ref{fig_model_1}.

\begin{figure}
\centering
\includegraphics[width=95mm]{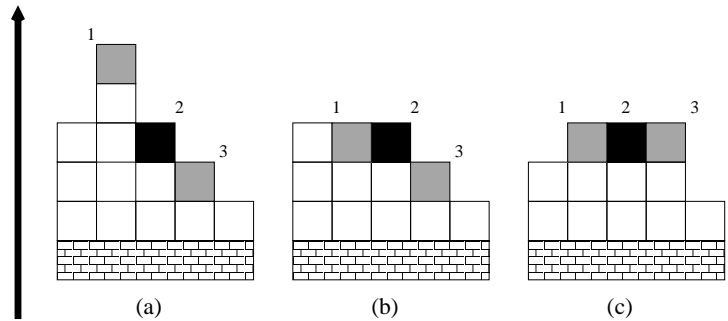}
\caption{Squares denote actually occupied sites in the lattice.
Gray sites denote possible places for incorporation to the surface, while the black square is the place of initial contact to the surface.
The line of bricks indicates substrate and fat arrow shows direction of growth.}
\label{fig_model_2}
\end{figure}

        So, the particle behavior after first contact to the surface depends on competition among values of $J$, $S$ and $V$.
(In calculations, we express them in $kT$ units.)
Negative values of $J$ correspond to attractive forces between particles and, therefore, a tendency to make the surface smoother.
Positive $J$ describes repulsion.
We assume negative values of $S$ ($S\le 0$).
The value of the diffusion barrier is always positive $(V\ge 0)$.
With increasing $V$, the probability of diffusion to places marked by arrows in Fig.~\ref{fig_model_1}c decreases.

	In Fig.~\ref{fig_model_2} black squares are places of initial impact of particle to surface.
The gray ones are possible places for particle incorporation after one step local relaxation.
A height difference in positions (1), (2) and (3) in Fig.~\ref{fig_model_2}a $(h_1>h_2>h_3)$ does not influence the preferences of the place of incorporation.
However, for positive values of $J$, place (1) is more attractive than (2) and (3) for minimum energy principle is then equivalent to minimum contact to neighbors.
And {\it vice versa}, for negative $J$  sites (2) or (3) are preferred as offering the strongest bond.
We must remember, however, that diffusion barrier $V$ or the edge $S$-term may alter the preferences.
In Fig.~\ref{fig_model_2}b both gray sites are equal energetically independently of the values of model control parameters.
For non-zero value of $S$ particles meet ES barrier when attempting position (3) in Fig.~\ref{fig_model_2}c.
In all three cases in Fig.~\ref{fig_model_2}, the increase in diffusion barrier $V$ promotes black positions (2).

\section{Results of Simulation}
\vspace*{-0.5pt}
\noindent
Our results of simulations for the surface morphology may be compared with the scaling law.
For a wide variety of discrete and continuum models of surface growth, the surface width $\sigma$---which describes surface roughness---is expected to follow the dynamical Family-Vicsek\citeup{family85,family86} scaling law:
\begin{subequations}
\label{eq_fv}
\begin{equation}
\sigma(t,L)\propto L^\alpha\cdot f(t/L^z)
\end{equation}
\begin{equation}
f(x)=
\begin{cases}
x^\beta \text{ and } z=\alpha/\beta & \text{ for } x \ll 1 \\
1                                   & \text{ for } x \gg 1 \\
\end{cases}
\end{equation}
\end{subequations}
where $\sigma^2=\langle h^2 \rangle -\langle h \rangle^2$ denotes the standard deviation of the surface height from the average surface height $\langle h \rangle$, $L$ is the linear size of the system, $t$ denotes time, the exponents $\alpha, \beta, z$ depend mainly on the system dimension, and $\langle\ldots\rangle$ denotes configurational average over all sites $\vec{r}$ at time $t$.

        Scaling versus both time and space, with two characteristic scaling exponents: roughness exponent $\alpha$ and dynamic exponent $z$, results from the universal self-affine scaling.
The surface profile and its properties are statistically invariant if the length in a direction parallel to the surface is scaled by a factor $\lambda$ and simultaneously the length in the perpendicular direction and the time $t$ by factors $\lambda^\alpha$ and $\lambda^z$, respectively.\citeup{herrmann86,gouyet91}
Such situation was also widely observed experimentally---see Ref. \cite{krim95} for review.

\subsection{Surface roughness}
\noindent
The simulations were carried out mainly on $500\times 500$ large square lattices with periodic boundary conditions.
The total number of ten monolayers were deposited.

	The scanning-tunneling microscopy (STM) studies\cite{linderoth97} show, that at higher temperatures particles can jump even over four lattice constants.
However single or double jumps are most frequent.
Therefore, and for simplicity of computations we decided to assume single step relaxation, to the nearest neighbors only $(L_{\text{dif}}=1)$.
The results of simulations of surface growth for different control parameters $J$, $S$ and $V$ are presented in Tables~\ref{tab_sigma_v_j}-\ref{tab_sigma_v_s}.

\subsubsection{Layer-by-layer growth}
\noindent
For negative values of binding energy $(J<0)$, and for the diffusion barrier small enough to allow for the migration, a layer-by-layer (or Frank-van der Merve) growth takes place, when each layer is completed before aggregation of island on higher levels.

\begin{figure}
\centering
\includegraphics[height=8cm,angle=-90]{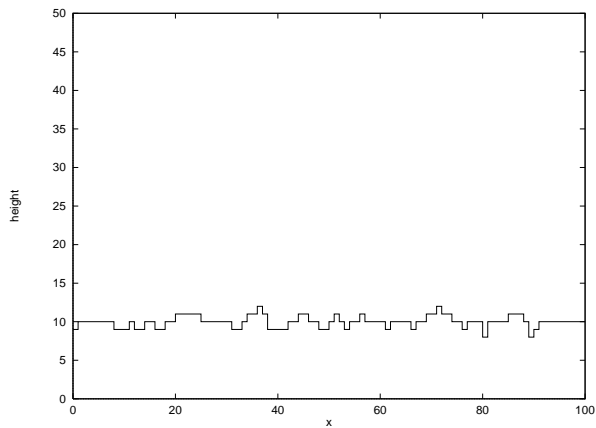}
\caption{Film profile for smooth surface ($J\to -\infty$, $\langle h \rangle$=10~ML).}
\label{fig_smooth}
\end{figure}

\begin{figure}
\centering
\includegraphics[height=8cm,angle=-90]{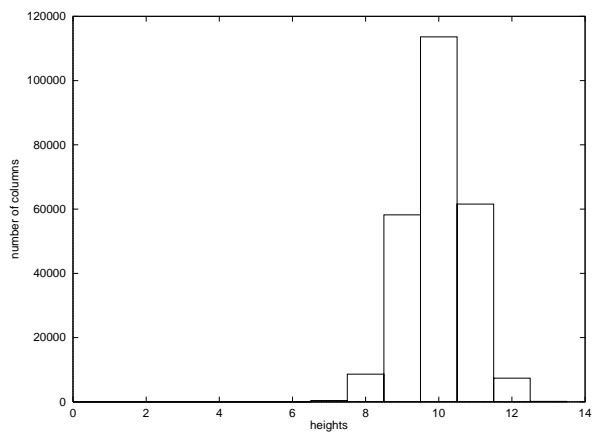}
\caption{Histogram of surface heights for smooth surface ($J\to -\infty$, $\langle h \rangle$=10~ML).}
\label{his_smooth}
\end{figure}

	In the limit $J\to -\infty$ a perfect wetting is expected for which $\sigma=0$, when all columns have the same heights $\langle h \rangle$.
However, from the simulation we got that for smooth surface limit (presented in Fig.~\ref{fig_smooth}) there are also the many sites with heights $\langle h\rangle\pm 1$ and some with $\langle h\rangle\pm 2$ (see histogram in Fig.~\ref{his_smooth}).
As result we get $\beta\approx 0.22$ instead of $\beta=0$.

	The deviation from ideally flat surface comes from limitation to one step only diffusion.
The limit $\sigma\to 0$ is for total energy minimum, while we seek local energy minimum only when employing local information on the site in question and its vicinity.
Then increasing the diffusion range to $L_{\text{dif}}$ is just $L_{\text{dif}}$ times repetition of the same local search.
This may either lead to the new final coordinates as far as $L_{\text{dif}}$ steps away from the initially hit site, yet still before the energy minimum is reached, or the particle may oscillate to and fro around the hit site if it happens to be the local minimum.
In each case, however, the final configuration is not the total energy minimum which would have been for the ideally flat surface.
Thus for the finite ranges $L_{\text{dif}}$ of diffusion, and for the applied algorithm of local search of energy minimum, we get in the $J\to-\infty$ limit $\sigma >0$.
The dependence of $\sigma(L_{\text{dif}},J\to-\infty)$ is discussed in chapter 3.1.5.
 
	Eq.~\eqref{eq_fv} shows that surface properties depends strongly on system size $L$.
Surface width $\sigma$ grows initially like $t^\beta$ until a steady state and saturated roughness $\sigma_\infty\propto L^\alpha$ after some time $\tau\propto L^z$ is reached.
Determination of the roughness exponent $\alpha$ is much more difficult for large lattices because a huge number of layers must be deposited before the
stationary state is reached, which requires enormous computer times.
We were able to measure roughness exponent $\alpha$ only for not too large lattices (max. $64\times 64$) and it seems that $\alpha\approx 0.78$.


\subsubsection{Random deposition}
\noindent
For $J=S=0$ (no driving force for particle migration) or for $V\to+\infty$ (when particles movements are frozen by the barrier), the surface growth corresponds to a simple random deposition model (``sit and stay'').
For $J=S=0$ we have
\[
p(\text{stay})=p(\text{left})=p(\text{right})=p(\text{forward})=p(\text{backward})=1/5,
\]
while in the latter case $V\to \infty$ we get
\[
p(\text{stay})=1 \qquad \text{and} \qquad p(\text{move})=0.
\]

\begin{figure}
\centering
\includegraphics[height=8cm,angle=-90]{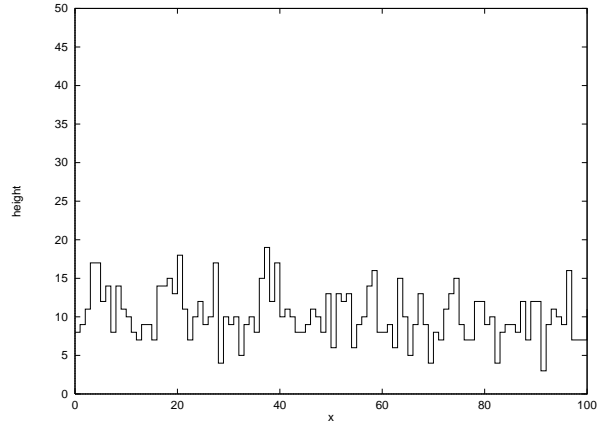}
\caption{Film profile for simple random deposition ($J=0$, $\langle h \rangle$=10~ML).}
\label{fig_random}
\end{figure}

\begin{figure}
\centering
\includegraphics[height=8cm,angle=-90]{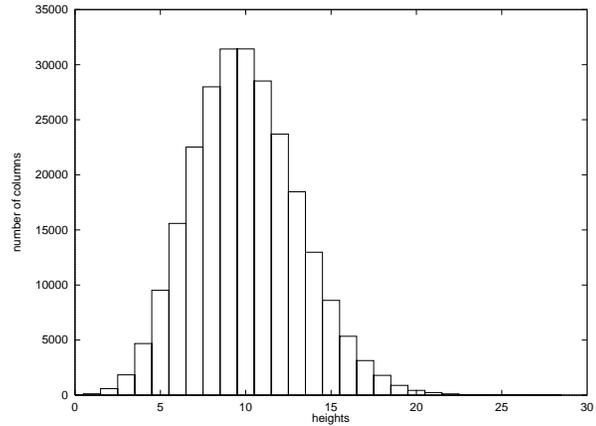}
\caption{Histogram of surface heights for random deposition ($J=0$, $\langle h \rangle$=10~ML)---Poisson distribution.}
\label{his_random}
\end{figure}

	The produced randomly surface profile is presented in Fig.~\ref{fig_random}.
We recovered the well known result, of the random deposition model that the column heights follow the Poisson distribution (see histogram in Fig.~\ref{his_random}).
That also implies $\sigma=\sqrt{t}$ $(\beta=1/2)$, the expected dependence of surface width $\sigma$ on time $t$ (film thickness $\langle h \rangle$).

\subsubsection{Growing of rough surfaces}
\noindent
When jumps to the top of higher neighbor columns are allowed, an unstable growth takes place and the surface width grows infinitely.\cite{kotrla92b}
For $J\to+\infty$ the dependence of surface width against time becomes linear.
Such rapid roughness growth $(\beta\approx 1)$ was experimentally observed by atomic force microscopy (AFM) studies of plasma-etched silicon surfaces\cite{brault98} and predicted by models of columnar growth.\cite{yao93}
Note, that this time the ``surface growth'' means removing some atoms instead of depositing them.

\begin{figure}
\centering
\includegraphics[height=8cm,angle=-90]{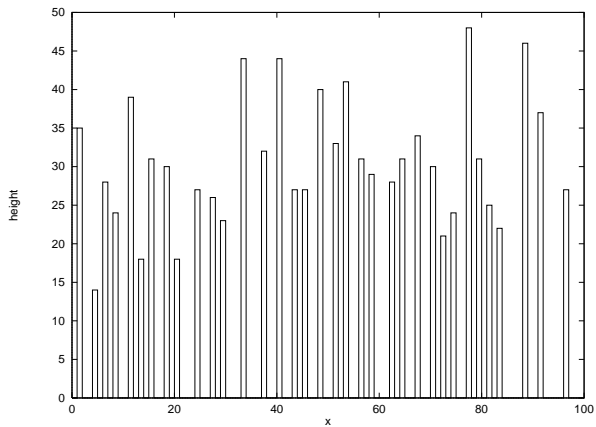}
\caption{Film profile for rough, spiky surface ($J\to+\infty$, $\langle h \rangle$=10~ML).}
\label{fig_rough}
\end{figure}

\begin{figure}
\centering
\includegraphics[height=8cm,angle=-90]{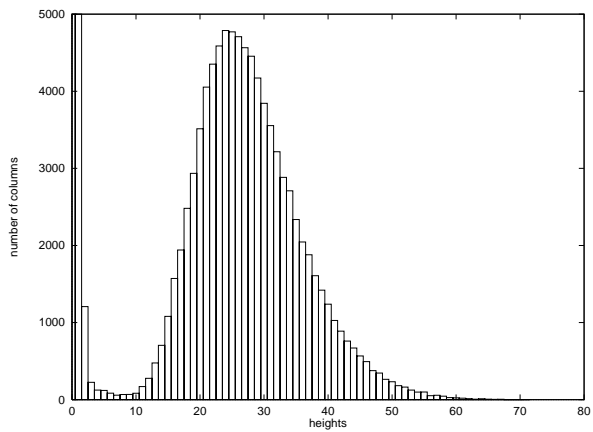}
\caption{Histogram of surface heights for rough surface ($J\to+\infty$, $\langle h \rangle$=10~ML).
$N(h=0)=160000$.}
\label{his_rough}
\end{figure}

	The surface profile for rough surface is presented in Fig.~\ref{fig_rough}.
In such case there are mainly unoccupied sites ($h=0$) while occupied sites
have two or three times larger heights than the
average surface height $\langle h \rangle$ (see Fig.~\ref{his_rough}).

%
%
\begin{table}
\centering
\caption{Surface width $\sigma$ [ML] for ten deposited layers, different values of diffusion barrier $V$ and interaction energy $J$, with $S=0$.}
\label{tab_sigma_v_j}
\begin{tabular}{r | rrrrrrr}
\hline
$V/J$
   &    -5&    -1& -0.5&    0&  0.5&      1&     5\\
\hline
  0&  0.87&  1.15& 1.52& 3.16& 7.34&  10.72& 14.17\\
  1&  0.88&  1.20& 1.58& 3.16& 7.59&  10.61& 13.90\\
  2&  0.88&  1.35& 1.80& 3.16& 6.94&   9.97& 13.70\\
  3&  0.89&  1.60& 2.17& 3.16& 5.45&   8.77& 13.44\\
  4&  0.92&  1.96& 2.58& 3.16& 4.16&   7.03& 13.13\\
  5&  0.96&  2.36& 2.88& 3.16& 3.52&   5.18& 12.71\\
 10&  1.39&  3.15& 3.16& 3.16& 3.17&   3.18& 11.29\\
 15&  1.98&  3.17& 3.16& 3.16& 3.16&   3.16&  9.29\\
 20&  2.61&  3.16& 3.17& 3.16& 3.16&   3.17&  5.57\\
 50&  3.16&  3.16& 3.16& 3.16& 3.16&   3.16&  3.17\\
\hline
\end{tabular}
\end{table}

%
%
\begin{table}
\centering
\caption{Surface width $\sigma$ [ML] for ten layers for different values of energies $J$ and $S$, with no diffusion barrier $V=0$.}
\label{tab_sigma_j_s}
\begin{tabular}{r | rrrrrrrrr}
\hline
$J/S$
     &    0&-0.25& -0.5&-0.75&   -1&   -2&   -3&   -4&   -5\\
\hline
    0& 3.16& 2.02& 1.55& 1.34& 1.23& 1.03& 0.97& 0.95& 0.95\\
-0.25& 2.02& 1.49& 1.28& 1.16& 1.09& 0.97& 0.93& 0.92& 0.92\\
 -0.5& 1.52& 1.25& 1.13& 1.06& 1.02& 0.94& 0.92& 0.91& 0.90\\
-0.75& 1.28& 1.13& 1.04& 0.99& 0.96& 0.91& 0.90& 0.90& 0.89\\
   -1& 1.15& 1.05& 0.99& 0.96& 0.94& 0.89& 0.89& 0.89& 0.89\\
   -2& 0.94& 0.92& 0.90& 0.89& 0.88& 0.87& 0.86& 0.87& 0.88\\
   -3& 0.90& 0.89& 0.88& 0.87& 0.86& 0.85& 0.86& 0.87& 0.87\\
   -4& 0.87& 0.87& 0.87& 0.86& 0.86& 0.85& 0.86& 0.86& 0.86\\
   -5& 0.88& 0.87& 0.87& 0.86& 0.87& 0.85& 0.85& 0.86& 0.86\\
\hline
\end{tabular}
\end{table}

%
%
\begin{table}
\centering
\caption{Surface width $\sigma$ [ML] for ten layers for different values of diffusion barrier $V$ and energy $S$, with $J=0$.}
\label{tab_sigma_v_s}
\begin{tabular}{r | rrrrrrrrr}
\hline
$V/S$
   &    0&-0.25& -0.5&-0.75&   -1&   -2&   -3&   -4&   -5\\
\hline
  0& 3.16& 2.02& 1.56& 1.34& 1.23& 1.03& 0.97& 0.95& 0.95\\
  1& 3.15& 2.07& 1.63& 1.40& 1.28& 1.06& 0.98& 0.96& 0.95\\
  5& 3.16& 3.04& 2.86& 2.62& 2.34& 1.62& 1.36& 1.22& 1.11\\
 10& 3.16& 3.16& 3.17& 3.15& 3.15& 2.82& 2.10& 1.76& 1.52\\
 50& 3.16& 3.16& 3.16& 3.16& 3.16& 3.16& 3.16& 3.16& 3.16\\
\hline
\end{tabular}
\end{table}

\subsubsection{Surface profile dynamics}
\noindent
The dynamical properties of film surface growth may be discussed in terms of growth exponent $\beta$ (or dynamic exponent $z$) and roughness exponent $\alpha$ in Eq.~\eqref{eq_fv}.
Numerically we found that growth exponent $\beta$ depends roughly only on the sign of interaction energy $J$ (when $S=V=0$):
\[
\beta=
\begin{cases}
0.22 & \text{ for } J\to -\infty\\
1/2 & \text{ for } J\to 0\\
1   & \text{ for } J\to +\infty\\
\end{cases}
\]
independently on lattice size $L$.

	Before completing the first monolayer a deviation from the true value of $\beta$ towards $1/2$ could be observed as an artifact of the
initially perfectly flat substrate.
Our results are consistent with some deterministic SOS models studied by Kotrla and Smilauer\cite{kotrla96a} (where $0.25 \le \beta \le 1$), Amar and Family\cite{amar98} SOS models for bcc structure ($\beta\approx 0.24$), and He {\it et al} experimental estimate\cite{he92} ($\beta\approx 0.22$).

\subsubsection{Diffusion range}
\noindent
So far presented results were produced for diffusion limited only to one inter-atomic separation ($L_{\text{dif}}=1$).
It was earlier suggested by Family\cite{family86} that growth mode is unchanged independently of diffusion range yet it is finite.
Here we check how diffusion range $L_{\text{dif}}$ influence on surface morphology for the case adequate to the Family model, it means $J\to-\infty$.
We found that an increasing diffusion range decreases surface roughness (see Table~\ref{tab_sigma_ldif_j}) until some saturation level $\sigma_{\text{sat}}$ is reached.
The saturation roughness level depends on binding energy $J$.
The random deposition case is recovered for $L_{\text{dif}}=0$ and/or $J=0$.

%
%
\begin{table}
\centering
\caption{Surface width $\sigma$ [ML] for ten layers for different one-step diffusion ranges $L_{\text{dif}}$ and binding energy $J$ ($V=0$ and $S=0$).
In all cases $\sigma(L_{\text{dif}})$ decreases with increasing $L_{\text{dif}}$ until some saturated roughness $\sigma_{\text{sat}}$ is reached.}
\label{tab_sigma_ldif_j}
\begin{tabular}{r | rrrrrrrrr}
\hline
$L_{\text{dif}}/J$&
          -20&    -5&    -1& -0.75&  -0.5& -0.25&  -0.1& -0.05&     0\\
\hline
0     &  3.16&  3.16&  3.16&  3.16&  3.16&  3.16&  3.16&  3.16&  3.16\\
1     &  0.87&  0.87&  1.15&  1.28&  1.52&  2.02&  2.61&  2.86&  3.16\\
2     &  0.82&  0.81&  1.03&  1.17&  1.42&  1.98&  2.61&  2.87&  3.16\\
5     &  0.80&  0.80&  0.94&  1.06&  1.29&  1.85&  2.54&  2.83&  3.16\\
50    &  0.68&  0.69&  0.93&  1.04&  1.26&  1.83&  2.52&  2.82&  3.16\\
500   &  0.47&  0.50&  1.00&  1.10&  1.30&  1.83&  2.53&  2.82&  3.16\\
5000  &  0.36&  0.49&  1.04&  1.11&  1.30&  1.83&  2.52&  2.84&  3.16\\
\hline
\end{tabular}
\end{table}

\subsubsection{Substrate temperature}
\noindent
Table~\ref{tab_sigma_temp} shows how the substrate temperature alters the surface roughness.
The values of model control parameters were taken after Ref.~11 
where simulation based on Arrhenius dynamics for homoepitaxy on Pt(111) substrate were carried out.
Here we identify our energetic diffusion barrier $V$ as the activation for diffusion energy ($V=0.75$~eV), the interaction energy $J$ as activation for breaking in-plane bonds between Pt atoms energy ($J=-0.15$~eV), and activation for breaking bonds with substrate energy ($S=-0.18$~eV) as our $S$.
The increase in temperature $T$ makes particles more mobile and then they may escape bonds easier, and across energetic barriers, which results in a decrease of surface roughness $\sigma$.
Such a situation is typical and well verified both experimentally and theoretically.\cite{smilauer93b}
However, no simple relation was reported for instance in Ref. \cite{liu97}.
%
%
\begin{table}
\centering
\caption{Surface width $\sigma$ after deposition of 20 layers for different substrate temperatures $T$ for homo-epitaxy growth on Pt(111) substrate: $V=0.75$~eV, $J=-0.18$~eV, $S=-0.15$~eV.}
\label{tab_sigma_temp}
\begin{tabular}{r rrrrr}
\hline
$T$~[K]  &  300&  600&  900& 1200& 1500\\
$\sigma$~[ML] & 2.04& 1.87& 1.74& 1.60& 1.49\\
\hline
\end{tabular}
\end{table}

\section{Conclusions}
\noindent
We obtain from the simulation $0.22\le\beta\le 1$ and that lower limit $\beta\approx 0.22$ is caused by the limited diffusions which pushes $\beta$ from $\beta=0$ (perfect wetting) towards $\beta=1/2$ (random deposition).

	We confirmed results from simulations based on deterministic rules of particles relaxation\cite{kotrla92b} that allowing or preventing particles to
climb to higher levels is essential for extracting the growth and roughness properties.
When such jumps are forbidden, then $\sigma$ grows accordingly to Family-Vicsek law~\eqref{eq_fv}.
In contrast, when jumps to the top of higher neighbor columns are
allowed, unstable growth takes place and surface width grows infinitely.

	Indeed, from the time evolution of surface profiles we confirm that for negative values of $J$ (when particles attract each other and as a result of that maximize the
number of bonds like in the Wolf-Villain model\cite{wolf90}---or minimize their heights like in the Family model\cite{family86}) a
stationary state is observed $(J\to-\infty, \alpha\approx 0.78, \beta\approx 0.22)$.
The saturation of roughness $(\lim_{t\to\infty}\sigma=\sigma_\infty)$
as a function of time (total coverage) was reported also experimentally for instance in Liu {\it et al} work\cite{liu97} for gold films growth on mica substrate under different experimental conditions.

	In the other limit, when particles repel each other and jump to higher levels (against gravitational force) the surface width growths infinitely and no stationary state is observed; this could be considered as infinite values of the roughness exponent $\alpha$ $(J\to +\infty, \alpha\to+\infty, \beta=1)$.

For finite ranges of diffusion, there exists some saturated roughness which does not allow for perfect wetting observed only in limiting case: 
\[
\lim_{J\to -\infty}\lim_{L_{\text{dif}}\to\infty}\sigma_{\text{sat}}=0.
\] 

\nonumsection{Acknowledgments}
\noindent
Main calculations were carried out in ACC-CYFRONET-AGH.
This work and machine in ACC-CYFRONET-AGH is supported by Polish Committee for Scientific Research (KBN) with grants no. 8~T11F~02616 and KBN/\-S2000/\-AGH/\-069/\-1998, respectively.

\nonumsection{References}


\begin{thebibliography}{20}

\bibitem{herrmann86}
H.~J.~Herrmann,
{\it Phys. Rep.} {\bf 136} (1986) 153.

\bibitem{gouyet91}
J.-F.~Gouyet, M.~Rosso, B.~Sapoval,
in {\it Fractals and Disordered Systems},
eds. A.~Bunde, S.~Havlin (Spinger-Verlag, Berlin, Heidelberg, 1991), p. 229.

\bibitem{levi97}
A.~C.~Levi, M.~Kotrla,
{\it J. Phys.: Condens. Matter} {\bf 9} (1997) 299.

\bibitem{kotrla92a}
M.~Kotrla,
{\it Czechoslovak J. Phys.} {\bf 42} (1992) 449.

\bibitem{maksymowicz96}
A.~Z.~Maksymowicz, M.~S.~Magdon, J.~S.~S.~Whiting,
{\it Comp. Phys. Comm.} {\bf 97} (1996) 101.

\bibitem{ehrlich66}
G.~Ehrlich, F.~G.~Hudda,
{\it J. Chem. Phys.} {\bf 44} (1966) 1039.

\bibitem{schwoebel66}
R.~L.~Schwoebel, E.~J.~Chipsey,
{\it J. Appl. Phys.} {\bf 37} (1966) 3682.

\bibitem{schwoebel69}
R.~L.~Schwoebel,
{\it J. Appl. Phys.} {\bf 40} (1969) 614.

\bibitem{markov96}
I.~Markov,
{\it Phys. Rev.} {\bf B54} (1996) 17930.

\bibitem{markov97}
I.~Markov,
{\it Phys. Rev.} {\bf B56} (1997) 12544.

\bibitem{smilauer93b}
P.~Smilauer, M.~R.~Wilby, D.~D.~Vvendesky,
{\it Phys. Rev.} {\bf B47} (1993) 4119.

\bibitem{family85}
F.~Family, T.~Vicsek,
{\it J. Phys.} {\bf A18} (1985) L75.

\bibitem{family86}
F.~Family,
{\it J. Phys.} {\bf A19} (1986) L441.

\bibitem{krim95}
J.~Krim, G.~Palasantzas,
{\it Int. J. Mod. Phys.} {\bf B9} (1995) 599.

\bibitem{linderoth97}
T.~R.~Linderoth, S.~Horch, E.~Laegsgaard, I.~Stensgaard, F.~Besenbacher,
{\it Phys. Rev. Lett.} {\bf 78} (1997) 4978.

\bibitem{kotrla92b}
M.~Kotrla, C.~Levi, P.~Smilauer,
{\it Europhys. Lett.} {\bf 20} (1992) 25.

\bibitem{brault98}
P.~Brault, P.~Dumas, F.~Salvan,
{\it J. Phys.: Condens. Matter} {\bf 10} (1998) L27.

\bibitem{yao93}
J.~H.~Yao, H.~Guo,
{\it Phys. Rev.} {\bf E47} (1993) 1007.

\bibitem{kotrla96a}
M.~Kotrla, P.~Smilauer,
{\it Phys. Rev.} {\bf B53} (1996) 13777.

\bibitem{amar98}
J.~G.~Amar, F.~Family,
{\it Surf. Rev. Lett.} {\bf 5} (1998) 851.

\bibitem{he92}
Y.~L.~He, H.~N.~Yang, T.~M.~Lu, G.~C.~Wang,
{\it Phys. Rev. Lett.} {\bf 69} (1992) 3770.

\bibitem{wolf90}
D.~E.~Wolf, J.~Villain,
{\it Europhys. Lett.} {\bf 13} (1990) 389.

\bibitem{liu97}
Z.~H.~Liu, N.~M.~D.~Brown, A.~McKinley,
{\it J. Phys.: Condens. Matter} {\bf 9} (1997) 59.

\end{thebibliography}
\end{document}